\documentclass[times, twocolumn]{aastex631}

\usepackage[flushleft]{threeparttable}
\usepackage{amsmath}
\usepackage{multirow}
\usepackage{bm}
\usepackage{pbox}
\usepackage{xcolor}

\shortauthors{Wang and Lin}

\begin{document}

\title{Uncalibrated Cosmic Standards as a Robust Test on Late-Time Cosmological Models}

\author{Yihao Wang}
\affiliation{South-Western Institute for Astronomy Research, Key Laboratory of Survey Science of Yunnan Province, Yunnan University, Kunming, Yunnan 650500, People's Republic of China}
\affiliation{Center for Joint Quantum Studies and Department of Physics, School of Science, Tianjin University, Tianjin 300350, China}

\author[0000-0003-2240-7031]{Weikang Lin}
\email{weikanglin@ynu.edu.cn}
\affiliation{South-Western Institute for Astronomy Research, Key Laboratory of Survey Science of Yunnan Province, Yunnan University, Kunming, Yunnan 650500, People's Republic of China}

\begin{abstract}
We present an assumption-minimized framework for testing late-time cosmological models using Uncalibrated Cosmic Standards (UCS)—including standard rulers and standard candles—without relying on absolute calibrations. The method exploits a tight, model-insensitive correlation between the sound horizons at recombination and the drag epoch. By avoiding dependence on pre-recombination physics and the amplitude of the Cosmic Microwave Background (CMB) power spectra, the UCS framework reduces potential early-Universe biases while retaining much of the constraining power of full analyses. Applying UCS to the recent dynamical dark energy (DE) study that reported deviations from $\Lambda$CDM \citep{DESI:2025zgx}, we find the constraints shift systematically toward the $\Lambda$CDM case. If this shift is physical, it may result from the omission of some pre-recombination physical processes that influence the scale dependence of the CMB spectra. We also observe a mild tension between uncalibrated standard rulers and candles, which can be largely mitigated by introducing a redshift-dependent magnitude bias in the supernova (SNe Ia) data. Our results highlight the importance of isolating post-recombination observables for testing late-time models in the era of precision cosmology, positioning UCS analysis as a robust framework for upcoming galaxy surveys.

\end{abstract}

\section{Introduction}\label{sec:intro}
The synergy of diverse cosmological observations has become a powerful tool for probing new physics—ranging from signals of the very early Universe \citep{Planck:2018jri}, to constraints on the sum of neutrino masses \citep{DESI:2025ejh}, to insights into dark sector properties \citep{Dong:2023jtk,DESI:2024mwx}, and beyond \citep{Abdalla:2022yfr}. However, joint analysis of different observations poses significant challenges. Each probe targets distinct astrophysical phenomena and involves different physical processes across cosmological epochs \citep{Lin_Mack_Hou_2020}, complicating systematic error control and result interpretation. The issue is particularly pressing, as tensions among cosmological observations or deviations from the standard cosmological model $\Lambda$CDM continue to emerge; see for instance \citep{Abdalla:2022yfr,DiValentino:2020zio,DiValentino:2020vvd,Dong:2023jtk,Riess:2024vfa,DESI:2024mwx,Vagnozzi:2019ezj,WL2017a,Lin:2019zdn,Gariazzo:2024sil}. 

To assess the robustness of the reported tensions or deviations, one effective strategy is to isolate and use only the observational information that is essential to the scientific objective, deliberately discarding information that only weakly improves the constraints. While this sacrifices some constraining power, it significantly reduces reliance on model-dependent assumptions and potential systematic errors, thereby enabling a more robust test of cosmological models. This is the goal of the present work, which focuses on cosmological scenarios involving novel post-recombination physics. 

Observations that map the history of cosmic expansion \citep{Sandage-1970-twonumbers} continue to serve as an effective test for late-time cosmological models \citep{Scolnic:2021amr,DESI:2024mwx}. Common probes include standard candles, such as SNe Ia, and standard rulers, such as the late-time Baryon Acoustic Oscillations (BAO). It is well-established that, even without calibration, these probes can provide strong constraints on late-time cosmological models \citep{SupernovaCosmologyProject:1996grv,Addison:2013haa}. 

\citet{Lin:2021sfs} introduced a unified framework to jointly use these ‘Uncalibrated Cosmic Standards (UCS)’ as a robust constraint on the matter density fraction ($\Omega_{\rm m}$) within the $\Lambda$CDM model. An important aspect of this framework is its incorporation of the measurement of sound horizon angular size at recombination ($\theta_*$) in a manner that is insensitive to early-Universe physics or potential systematics associated with the amplitude of the CMB power spectra.\footnote{Earlier studies employed an uninformative prior on the baryon density, effectively leaving the sound horizon unconstrained \citep{2015-Aubourg-BOSS-collaboration}, but the approach developed in \citet{Lin:2021sfs} more clearly demonstrates the independence from early-Universe physics.} The key idea is that while the sound horizon scale is influenced by early-Universe physics \citep{Knox:2019rjx,Aylor:2018drw,Poulin:2018cxd}, the (normalized) difference between the sound horizon at recombination and at the drag epoch is insensitive to conditions prior to recombination. We dub the framework the UCS analysis.

Several studies have tested cosmological models using various combinations of observational datasets or subsets of data \citep{Giare:2025pzu,Colgain:2025nzf,Gialamas:2024lyw}. In contrast, the UCS analysis isolates the most relevant information for late-time cosmology, minimizing underlying assumptions at the cost of only mildly reduced constraining power. 

In this work, we extend the UCS analysis to test several widely used late-time cosmological models, with a particular focus on dynamical DE as described by the CPL parameterization \citep{Chevallier:2000qy,Linder:2002et}, about which studies have reported significant deviations from the $\Lambda$CDM model \citep{DESI:2024hhd,DESI:2025zgx,DESI:2025fii}. 

We adopt geometric units where $G=c=1$. Our code for UCS analysis is publicly available at \href{https://github.com/WeikangLin/UCS-Late-Cosmology.git}{https://github.com/WeikangLin/UCS-Late-Cosmology.git}.

\section{Method}\label{sec:method}
\subsection{Key assumption in the UCS analysis}
The key to the UCS analysis is the tight correlation between the sound horizons at recombination and at the drag epoch. More explicitly, the smallness and the value of the following Hubble-constant-normalized difference in the sound horizon is rather insensitive to alternative cosmological models:
\begin{equation}\label{eq:difference-rH0}
    \Delta rH_0\equiv(r_{\rm d}-r_\ast)H_0=\int^{z_*}_{z_d}\frac{c_{\rm{s}}(z)}{E(z)}dz\,,
\end{equation}
The UCS analysis takes advantage of the fact that $\Delta rH_0$ is small compared to $r_\ast H_0$ due to the narrow gap in redshift between recombination and the drag epoch. As long as this condition holds, UCS-derived constraints only weakly depend on the exact value of $\Delta rH_0$. For instance, even a doubling of $\Delta rH_0$—due to unknown but extreme non-standard physics—would change the inferred $\Omega_{\rm m}$ in $\Lambda$CDM by only about $3.6\%$ \citep{Lin:2021sfs}. Moreover, the value of $\Delta rH_0$ is inherently difficult to modify significantly; achieving a substantial change would require some drastic deviations from $\Lambda$CDM, particularly between recombination and the drag epoch \citep{Lin:2021sfs}.

Further, while the exact value of $\Delta rH_0$ does slightly vary in different models, it has already been captured via Eq.~\eqref{eq:difference-rH0}. For a range of non-standard post-recombination models (such as DE), the small variations in $\Delta rH_0$ primarily arise from changes in the resulting $\Omega_{\rm m}$, since $E \simeq \sqrt{\Omega_{\rm m}/a^3}$ between $z_{\rm d}$ and $z_*$. Notably, the quantity $\Delta rH_0\sqrt{\Omega_{\rm m}}$ remains remarkably stable across a broad class of non-standard models. Accordingly, in the fiducial UCS analysis, we adopt
\begin{equation}\label{eq:UCS-simplified-prior}
\sqrt{\Omega_{\rm m}}\Delta rH_0 = (3.36 \pm 0.7) \times 10^{-4}\,,
\end{equation}
as inferred from the MCMC chains provided by Planck assuming the $\Lambda$CDM model \citep{2018-Planck-cosmo-params}. The constraint on $\sqrt{\Omega_{\rm m}}\Delta rH_0$ is nearly the same across the models investigated in \citet{2018-Planck-cosmo-params}, such as the CPL DE model and the curved-$\Lambda$CDM, with only some slight changes in the mean and error bar.

In scenarios where additional energy injection occurs between recombination and the drag epoch, or where the meaning of $\Omega_{\rm m}$ varies substantially from recombination to the present day, Eq.\,\eqref{eq:difference-rH0} may be used to compute $\Delta r H_0$ explicitly, rather than relying on the simplified form given by Eq.\,\eqref{eq:UCS-simplified-prior}; see \citet{Lin:2021sfs} for details. Nevertheless, \textit{a foundational assumption of the UCS framework is that models of interest do not substantially alter $\Delta r H_0$, so that $\Delta r H_0 \ll r_{\rm d} H_0$ remains valid. This is a central and fairly robust assumption.}

\subsection{Key quantities in the cosmic expansion history}\label{sec:background-quantities}
The most general cosmological scenario considered in this work is a spatially curved FLRW universe containing a dynamical DE component parameterized by the CPL form,  $w(a) = w_0 + (a - 1)w_a$, along with pressureless matter (cold dark matter and baryons), photons, and one massless and two massive neutrinos in a normal mass hierarchy. The key observable for probes of the cosmic expansion history is the transverse comoving distance in units of Hubble distance as a function of redshift:
\begin{equation}\label{eq:fM-z}
    H_0D_{\textsc{m}}(z)=\frac{1}{\Omega_{\textsc{k}}^{1/2}}\sinh\Big[\Omega_{\textsc{k}}^{1/2}\int_0^z\frac{dz'}{E(z')}\Big]\,,
\end{equation}
where we take $\sqrt{-1}=i$ and $E(z)$ is given by
\begin{equation}\label{eq:Ez}
    E(z)=\sqrt{\Omega_{\textsc{de}}e_{\textsc{de}}(z)+\frac{\Omega_{\textsc{k}}}{a^2}+\frac{\Omega_{\rm m}}{a^3}+\frac{\Omega_{\rm r}}{a^4}+\sum_i\Omega_{\nu_i}e_i(z)}\,.
\end{equation}
In this expression, the $\Omega$'s denote the present-day fractional energy densities of dark energy, spatial curvature, matter, radiation, and massive neutrinos, respectively. With the CPL parameterization, it can be shown that
\begin{equation}\label{eq:DE-energy-fraction-wwa}
    e_{\textsc{de}}(z)=(1+z)^{3(1+w_0+w_a)}\exp\big(-\frac{3w_az}{1+z}\big)\,.
\end{equation}
The radiation and massive neutrino components contribute only marginally to $E(z)$ and $H_0D_{\textsc{m}}(z)$. We include them for completeness, although their impact on the analysis is negligible. Assuming the thermal relic scenario with one massless neutrino and a normal mass hierarchy we adopt $\Omega_{\rm{r}}=\Omega_{\gamma}+\Omega_{\nu}^{\rm{massless}}=3.0337\,h^{-2}\times10^{-5}$ and $\Omega_{\nu_i}=\frac{ m_{\nu_i}}{93.14\,h^2\, {\rm eV}}$ \citep{2012-Lesgourgues-Pastor-nu-mass}
with neutrino masses $m_{\nu_1} = 0.0087,\text{eV}$ and $m_{\nu_2} = 0.0494,\text{eV}$, as inferred from the mass-squared differences obtained from neutrino oscillation \citep{Olive_2014}. The normalized redshift evolution of the energy density for each massive neutrino is approximated by \citep{Lin:2023fao}
\begin{equation}\label{eq:evolution-nu}
    e_i(z)=\frac{1}{a^4}\left(\frac{a^n+a_{{\rm{T}},i}^n}{1+a_{{\rm{T}},i}^n}\right)^{\frac{1}{n}}\,,
\end{equation}
with $n=1.8367$ and $a_{{\rm{T}},i}=3.1515\frac{T_\nu^0}{m_{\nu_i}}$. The closure of the Universe's energy budget imposes,
\begin{equation}\label{eq:energy-budget-closure}
    \Omega_{\textsc{de}}+\Omega_{\textsc{k}}+\Omega_{\rm m}+\Omega_{\rm r}+\sum_i\Omega_{\nu_i}=1\,.
\end{equation}

The priors of parameters are listed in Table~\ref{tab:parameter-prior-UCS}.

\begin{table}[tbp]
    \centering
    \caption{Priors on the parameters used in this work. A uniform prior on a quantity q is denoted as $\mathcal{U}[q_{\rm min},\ q_{\rm max}]$, and a Gaussian prior as $\mathcal{G}(\bar{q},\ \sigma_q)$. The last row represents a key assumption in the fiducial UCS analysis, which remains robust across a wide range of cosmological models.}
    \label{tab:parameter-prior-UCS}
    \begin{tabular}{l @{\hspace{1.5cm}} l}
    \hline
    Parameter & Prior \\
    \hline
    $\Omega_{\rm m}$ & $\mathcal{U}[0.01,\ 0.99]$ \\
    $w_0$ & $\mathcal{U}[-3,\ 1]$ \\
    $w_a$ & $\mathcal{U}[-3,\ 2]$ \\
    $\Omega_{\rm K}$ & $\mathcal{U}[-0.3,\ 0.3]$ \\
    $r_{\rm d}H_0$ & $\mathcal{U}[0.01,\ 0.1]$ \\
    $\mathcal{M}$ & $\mathcal{U}[17,\ 30]$ \\
    \hline
    $\sqrt{\Omega_{\rm M}}\Delta rH_0$ & $\mathcal{G}(0.000336,\ 0.00007)$  \\
    \hline
    \end{tabular}
\end{table}

\subsection{Standard rulers}\label{sec:std-rulers}
Standard rulers, such as the sound horizon scales at recombination and the drag epoch, have fixed comoving sizes. Their observables are,
\begin{align}
D_{\rm M}/r_{\rm d} &= \frac{H_0 D_{\textsc{m}}(z_{\rm eff})}{r_{\rm d} H_0}\,, \label{eq:angular-size-BAO} \\
\theta_* &= \frac{r_* H_0}{H_0 D_{\textsc{m}}(z_*)}\,, \label{eq:angular-size-CMB}
\end{align}
and 
\begin{equation} \label{eq:zspan-BAO}
D_H/r_{\rm d} = \frac{1}{r_{\rm d} H_0 E(z_{\rm eff})}\,,
\end{equation}
all evaluated at their respective effective redshifts. We multiply both the numerator and the denominator with $H_0$ in Eqs.\,\eqref{eq:angular-size-BAO} and \eqref{eq:angular-size-CMB} in order to explicitly separate the free parameter $r_{\rm d} H_0$ and normalized comoving distance $H_0D_{\textsc{m}}$, the latter of which is independent of $H_0$ as evident from the right-hand side of Eq.\,\eqref{eq:fM-z}.\footnote{We note that, to unify the description of CMB and late-time BAO measurements, \citet{Lin:2021sfs} adopted the notation of \citet{Seo:2003pu} and express the BAO observables in terms of the angular size of the standard ruler when placed perpendicular to the line of sight, $\theta_{r_{\rm d}}$, and the redshift separation when aligned along the line of sight, $\Delta z_{r_{\rm d}}$. These quantities are simply the reciprocals of the present-day more commonly reported ratios $D_{\rm M}/r_{\rm d}$ and $D_{\rm H}/r_{\rm d}$, respectively.}

Since only dimensionless quantities are directly observed, the comoving horizon sizes are degenerate with the Hubble constant unless an early-Universe cosmological model is assumed \citep{Addison:2013haa}. Consequently, the product $r_{\rm d} H_0$ is treated as a free parameter. \textit{The strong correlation between two sound horizon scales allows us to treat $r_\ast H_0$ as a derived parameter, given $r_{\rm d} H_0$ and Eq.~\eqref{eq:UCS-simplified-prior}.}

The uncalibrated standard-ruler likelihood is given by
\begin{equation}\label{eq:bao-likelihood}
\begin{split}
    \ln \mathcal{L}_{\rm{\textsc{ucs}}}^{\rm ruler}=&\sum\limits_i-\frac{1}{2}(\bm{\Theta_i}-\bm{Q_i})^T\bm{C_i}^{-1}(\bm{\Theta_i}-\bm{Q_i})\\
      &-\frac{(\theta_*^{\rm pre.}-\theta_*^{\rm meas.})^2}{2\sigma_{\theta_*}^2}+\rm{Const.}
\end{split}
\end{equation}
Here, $\bm{\Theta_i}$ and $\bm{Q_i}$ denote the predicted and measured values of $(\theta_{\rm{d}},~\Delta z_{r_{\rm d}})$, or their combined quantity $D_{\rm V}/r_{\rm d}=(z_{\rm eff}D_{\rm M}^2D_{\rm H})^{1/3}/r_{\rm d}$ when they are not reported individually.
The matrix $\bm{C_i}$ is the covariance matrix corresponding to the $i$th effective redshift. The terms $\theta_*^{\rm pre.}$ and $\theta_*^{\rm meas.}$ represent the predicted and measured values of $\theta_*$. The standard-ruler measurements used in this analysis are obtained from the latest DESI BAO results \cite{DESI:2025zgx}, labeled as DESI2025, as well as the angular size of the sound horizon at recombination, $\theta_*$, taken from \citet{2018-Planck-cosmo-params}. We use $+\theta_*^{\textsc{ucs}}$ to denote the UCS analysis, and $+\theta_*^{\rm Early}$ to denote the conventional analysis that relies on an early-Universe modeling of the sound horizon scale and includes the strong CMB priors on $\Omega_{\rm m}h^2$ and $\Omega_{\rm b}h^2$.

While uncalibrated SNe Ia and uncalibrated late-time BAO are well-established methods of analysis, it is important to highlight the role of the uncalibrated CMB sound horizon scale. Conventional joint analyses involving the CMB—whether through the full CMB likelihood, compressed constraints on $\theta_*$ and the shift parameter \citep{Bond:1997wr,Efstathiou:1998xx}, or compressed constraints on ($\theta_*$, $\Omega_{\rm m}h^2$, $\Omega_{\rm b}h^2$) \citep{Lemos:2023xhs}—explicitly rely on early-Universe cosmological models and information derived from the amplitude of CMB power spectra, thereby ``calibrating'' $r_*$ \citep{Verde:2019ivm}. In such analyses, the absolute scale of the sound horizon at recombination is computed assuming a specific early-Universe model. This introduces a potential model-dependent bias into the inference of late-time parameters, as the sound horizon scale correlates with them (not only $H_0$). Moreover, the scale dependence of the CMB spectrum amplitude imposes strong priors on parameters like $\Omega_{\rm m}h^2$, regardless of whether the full or compressed analysis is used. This could introduce bias in late-time parameters from systematic errors related to the CMB spectrum amplitude.

In contrast, the UCS analysis improves both aspects discussed above in a more model-independent manner.

First, UCS does not assume any specific early-Universe cosmological model. Instead, it relies on the tight, model-insensitive correlation between the two sound horizon scales. As shown in Eq.\,\eqref{eq:angular-size-CMB}, $r_*H_0$ enters as a parameter when incorporating the measurement of $\theta_*$. However, treating $r_*H_0$ as an independent free parameter offers little cosmological insight, as its value would simply be inferred from Eq.\,\eqref{eq:angular-size-CMB} via the late-time cosmological parameters constrained by BAO and SNe Ia. Rather than adopting this uninformative approach, UCS leverages the robust correlation between $r_*H_0$ and $r_{\rm d}H_0$, allowing the two to be treated as nearly equivalent, differing only by the small offset specified in Eq.\,\eqref{eq:difference-rH0} or its simplified form in Eq.\,\eqref{eq:UCS-simplified-prior}. As a result, both $r_*H_0$ and $r_{\rm d}H_0$ are determined jointly from observations, rather than being derived from an early-Universe model.

Second, UCS uses only the purely geometrical information—the measurement of the sound horizon angular scale $\theta_*$ from CMB, deliberately excluding any information related to the amplitude of CMB fluctuations. This selective use further reduces reliance on early-Universe assumptions and avoids potential systematic uncertainties associated with the CMB spectrum amplitude. Despite excluding amplitude-related information, the inclusion of the tightly constrained $\theta_*$ still provides a powerful constraint on late-time cosmological parameters.

\begin{figure*}
    \centering
    \includegraphics[width=0.245\textwidth,height=0.245\textwidth]{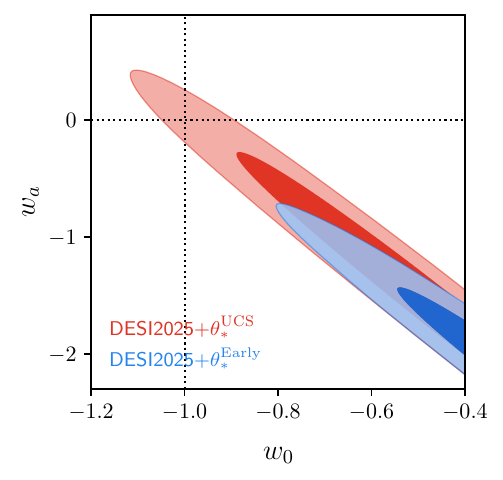}
    \includegraphics[width=0.245\textwidth,height=0.245\textwidth]{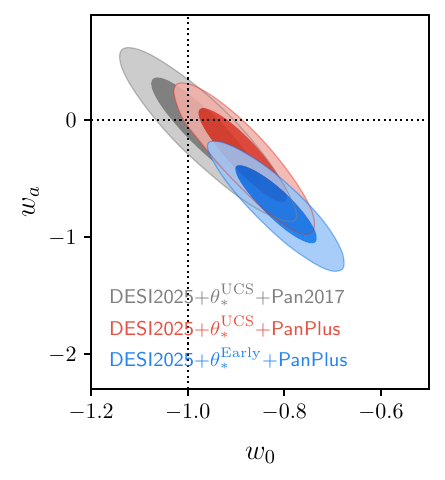}
    \includegraphics[width=0.245\textwidth,height=0.245\textwidth]{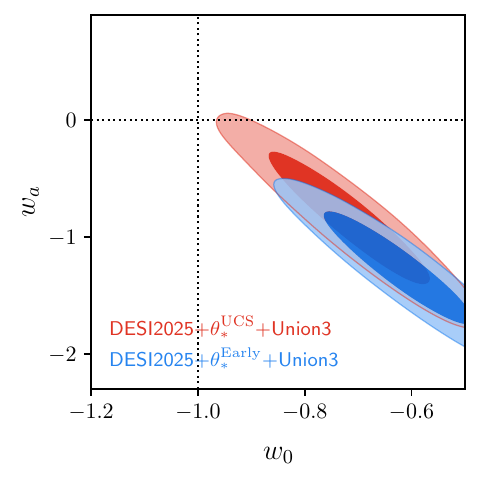}
    \includegraphics[width=0.245\textwidth,height=0.245\textwidth]{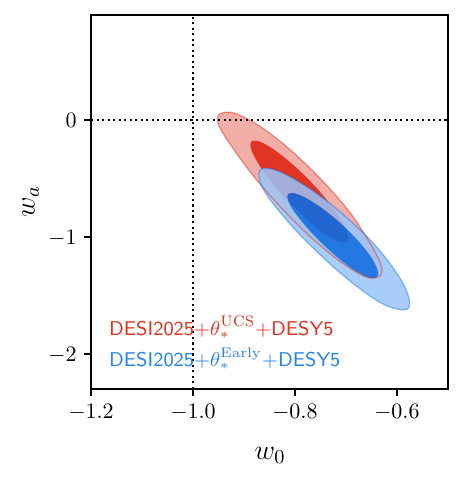}
    \caption{Comparison of constraints on the $w_0$–$w_a$ plane. We combine BAO measurements from DESI 2025 \citep{DESI:2025zgx}, the angular acoustic scale from Planck \citep{2018-Planck-cosmo-params}, and SNe Ia data from multiple compilations. First: without SNe Ia data; Second: with PanPlus \citep{Scolnic:2021amr} and Pan2017 \citep{Scolnic-etal-2018-Pantheon}; Third: with Union3 \citep{Rubin:2023ovl}; Fourth: with DESY5 \citep{DES:2024jxu}. In all panels, blue contours (CMB prior) reproduce results reported from \citet{DESI:2025zgx}, while red contours show constraints from the UCS analyses. By removing assumptions of pre-recombination physics and potential systematics associated with the CMB power spectra amplitude, the UCS analyses yield results more consistent with $\Lambda$CDM. In the second panel, gray contours represent constraints using the earlier compilation Pan2017 which–despite a somewhat weaker constraining power with an older analysis method–remain in excellent agreement with $\Lambda$CDM.}
    \label{fig:w0-wa-compare}
\end{figure*}

\subsection{Standard candles}\label{sec:std-candles}
Standard candles are astronomical objects with constant intrinsic luminosity. While the analysis of uncalibrated SNe Ia is well established in the literature, we briefly outline the key steps here for completeness. For SNe Ia, their intrinsic peak luminosity at some reference stretch and reference color is assumed to be constant \citep{1998A&A...331..815T}. The B-band apparent magnitude of SNe Ia is given by \citep{SupernovaCosmologyProject:1996grv,Scolnic:2021amr}:
\begin{equation}\label{eq:magnitude-redshift}
m_{\textsc{b}} = 5\log_{10}\big[(1+z_{\textsc{hel}})H_0 D_{\textsc{m}}(z_{\textsc{hd}}) \big]+\mathcal{M}\,,
\end{equation}
where $m_{\textsc{b}}$ is the systematic-corrected B-band apparent magnitude, $z_{\textsc{hel}}$ and $z_{\textsc{hd}}$ are the heliocentric redshift and Hubble diagram redshift, respectively, and $\mathcal{M}$ is a combination of the Hubble constant and peak absolute magnitude $M_0$ of the SNe Ia at the reference stretch and color \citep{Lin:2021sfs}:
\begin{equation}\label{eq:C-H-M}
\mathcal{M}\equiv M_0 - 5\log_{10}(10\,{\rm{pc}}\times H_0)\,.
\end{equation}
As seen in Eq.~\eqref{eq:magnitude-redshift}, $M_0$ and $H_0$ are degenerate, making $\mathcal{M}$ a free parameter unless $M_0$ is independently calibrated using primary distance indicators \citep{Goliath:2001af}.

The SNe Ia likelihood is given by 
\begin{equation}\label{eq:sn-likelihood}
 \ln\mathcal{L}_{\textsc{sn}}=
   -\frac{1}{2}(\bm{m_{\textsc{b}}^{\rm{o}}}-\bm{m_{\textsc{b}}^{\rm{p}}})^{\rm{T}}\bm{C}^{-1} 
  (\bm{m_{\textsc{b}}^{\rm{o}}}-\bm{m_{\textsc{b}}^{\rm{p}}})+\rm{Const.}    \,,
\end{equation}
where $\bm{m_{\textsc{b}}^{\rm{o}}}$ and $\bm{m_{\textsc{b}}^{\rm{p}}}$ are the observed and predicted B-band apparent magnitudes at reference stretch and color, respectively, and $\bm{C}$ is the covariance matrix that includes both statistical and systematic uncertainties. We use four SNe Ia compilations, including the latest Pantheon+ SN analysis (PanPlus) \citep{Scolnic:2021amr}, and earlier Pantheon compilation (Pan2017) \citep{Scolnic-etal-2018-Pantheon}, the lastest Union compilation (Union3) \citep{Rubin:2023ovl} and the full five years of the Dark Energy Survey (DESY5) \citep{DES:2024jxu}. For the Union3 dataset, where only the estimated distance moduli at certain effective redshifts are provided, we compute the inferred apparent magnitudes $m_{\textsc{b}}$ by adding a reference absolute magnitude $M_0 = -19.3$ to the distance moduli. These inferred magnitudes correspond to the apparent brightness that Type Ia supernovae (SNe Ia) with reference stretch and color would exhibit if located at the given redshifts.

\section{Result}\label{sec:results}

We begin by closely reproducing the results presented in Figure 11 of \citet{DESI:2025zgx} for the DE equation of state parameters using the CPL parameterization in a spatially flat Universe. These are shown as the blue contours in Figure~\ref{fig:w0-wa-compare}. Instead of employing the full CMB likelihood, we adopt the compressed constraints on \mbox{($\theta_*$, $\Omega_{\rm m}h^2$, and $\Omega_{\rm b}h^2$)} from \citet{Lemos:2023xhs}, which were shown in \citet{DESI:2025zgx} to retain the most information in the full analysis for constraining late-time models.

The red contours in Figure~\ref{fig:w0-wa-compare} depict the parameter constraints obtained from our UCS analyses. Although the UCS analysis results in a modest degradation of constraining power, there is a notable and systematic shift of the parameter space toward the standard $\Lambda$CDM baseline $(w_0 = -1, w_a = 0)$. Importantly, all UCS results remain consistent with the $\Lambda$CDM model within, or marginally beyond, the 2-$\sigma$ confidence region. 

We emphasize that the degradation in constraining power from the UCS analysis is quite modest, as it already retains the most relevant information for constraining late-time cosmological models. To quantify this, we compute the Figures of Merit (FoMs) for the constraints in the $w_0$–$w_a$ plane \citep{Wang:2008zh}:
\begin{equation}\label{eq:FOM}
{\rm FoM}_{w_0,w_a}=\frac{1}{\sqrt{{\rm det~Cov}(w_0,w_a)}},.
\end{equation}
For the constraints shown in the second to fourth panels of Figure~\ref{fig:w0-wa-compare}, the FoMs from the UCS analyses (red) decrease by only a factor of 1.07 to 1.16 relative to the conventional analyses (blue).

In the second panel of Figure~\ref{fig:w0-wa-compare}, we also show the results (gray contours) obtained using the earlier Pantheon compilation \citep{Scolnic-etal-2018-Pantheon}, instead of the latest release \citep{Scolnic:2021amr}. Interestingly, the resulting constraints are fully consistent with the $\Lambda$CDM model.

Table~\ref{tab:summary-results} presents the median and the 1-$\sigma$ upper and lower limits for each parameter, marginalized over the other parameters for the late-time cosmological models considered. Compared to the results obtained from the corresponding full analysis, the minimal set of assumptions adopted in the UCS analysis enhances the robustness of the results.

\begin{table*}[htbp]
    \begin{ruledtabular}
    \footnotesize
    \renewcommand{\arraystretch}{1.15}
    \caption{Summary statistics of the UCS analysis results for the late-time cosmological models considered in this work. For each parameter, the median and the 1-$\sigma$ upper and lower bounds marginalized over the other parameters are listed. \label{tab:summary-results}}
    \begin{tabular}{llcccccc}
        Model & Data (UCS analysis) & $\Omega_{\rm m}$  & $w_0$ & $w_a$ & $\Omega_{\rm K}\times10^3$ & $r_{\rm d}H_0$ & $\mathcal{M}$\\
        \hline
        $\Lambda$CDM & DESI BAO+$\theta_*^{\textsc{ucs}}$ & ${0.2960}^{+0.0049}_{-0.0049}$ & -- & -- & -- & ${0.03386}^{+0.00019}_{-0.00019}$ & -- \\
        & DESI BAO+$\theta_*^{\textsc{ucs}}$+PanPlus & ${0.2985}^{+0.0045}_{-0.0047}$ & -- & -- & -- & ${0.03376}^{+0.00019}_{-0.00018}$ & ${23.7969}^{+0.0040}_{-0.0040}$ \\
        & DESI BAO+$\theta_*^{\textsc{ucs}}$+Union3 & ${0.2969}^{+0.0042}_{-0.0041}$ & -- & -- & -- & ${0.03380}^{+0.00018}_{-0.00019}$ & ${23.7828}^{+0.0890}_{-0.0897}$ \\
        & DESI BAO+$\theta_*^{\textsc{ucs}}$+DESY5 & ${0.3006}^{+0.0046}_{-0.0046}$ & -- & -- & -- & ${0.03368}^{+0.00019}_{-0.00018}$ & ${23.8259}^{+0.0051}_{-0.0049}$ \\
        \hline
        $w$CDM & DESI BAO+$\theta_*^{\textsc{ucs}}$ & ${0.2998}^{+0.0076}_{-0.0077}$ & ${-0.969}^{+0.045}_{-0.047}$ & -- & -- & ${0.03357}^{+0.00047}_{-0.00044}$ & --\\
        & DESI BAO+$\theta_*^{\textsc{ucs}}$+PanPlus & ${0.3032}^{+0.0054}_{-0.0054}$ & ${-0.946}^{+0.027}_{-0.028}$ & -- & -- & ${0.03335}^{+0.00028}_{-0.00027}$ & ${23.8074}^{+0.0066}_{-0.0067}$ \\
        & DESI BAO+$\theta_*^{\textsc{ucs}}$+Union3 & ${0.3063}^{+0.0057}_{-0.0060}$ & ${-0.926}^{+0.034}_{-0.033}$ & -- & -- & ${0.03315}^{+0.00034}_{-0.00034}$ & ${23.7895}^{+0.0922}_{-0.0899}$ \\
        & DESI BAO+$\theta_*^{\textsc{ucs}}$+DESY5 & ${0.3075}^{+0.0055}_{-0.0054}$ & ${-0.917}^{+0.025}_{-0.026}$ & -- & -- & ${0.03307}^{+0.00027}_{-0.00026}$ & ${23.8536}^{+0.0101}_{-0.0100}$ \\
        \hline
        $w_0w_a$CDM & DESI BAO+$\theta_*^{\textsc{ucs}}$ & ${0.3470}^{+0.0260}_{-0.0290}$ & ${-0.486}^{+0.261}_{-0.291}$ & ${-1.534}^{+0.908}_{-0.840}$ & -- & ${0.03149}^{+0.00127}_{-0.00103}$ & -- \\
        & DESI BAO+$\theta_*^{\textsc{ucs}}$+PanPlus & ${0.3075}^{+0.0066}_{-0.0066}$ & ${-0.884}^{+0.060}_{-0.058}$ & ${-0.319}^{+0.266}_{-0.270}$ & -- & ${0.03325}^{+0.00030}_{-0.00029}$ & ${23.8144}^{+0.0089}_{-0.0089}$ \\
        & DESI BAO+$\theta_*^{\textsc{ucs}}$+Union3 & ${0.3226}^{+0.0100}_{-0.0102}$ & ${-0.719}^{+0.104}_{-0.099}$ & ${-0.830}^{+0.390}_{-0.391}$ & -- & ${0.03250}^{+0.00045}_{-0.00043}$ & ${23.8087}^{+0.0893}_{-0.0936}$ \\
        & DESI BAO+$\theta_*^{\textsc{ucs}}$+DESY5 & ${0.3164}^{+0.0066}_{-0.0068}$ & ${-0.790}^{+0.067}_{-0.064}$ & ${-0.638}^{+0.297}_{-0.305}$ & -- & ${0.03286}^{+0.00029}_{-0.00028}$ & ${23.8731}^{+0.0137}_{-0.0135}$ \\
        \hline
        $\Omega_{\rm K}\Lambda$CDM & DESI BAO+$\theta_*^{\textsc{ucs}}$ & ${0.2972}^{+0.0079}_{-0.0079}$ & -- & -- & ${0.4}^{+3.2}_{-3.1}$ & ${0.03382}^{+0.00025}_{-0.00024}$ & -- \\
        & DESI BAO+$\theta_*^{\textsc{ucs}}$+PanPlus & ${0.3033}^{+0.0074}_{-0.0072}$ & -- & -- & ${2.6}^{+3.1}_{-2.9}$ & ${0.03365}^{+0.00023}_{-0.00023}$ & ${23.7990}^{+0.0046}_{-0.0048}$ \\
        & DESI BAO+$\theta_*^{\textsc{ucs}}$+Union3 & ${0.3025}^{+0.0077}_{-0.0075}$ & -- & -- & ${2.6}^{+3.1}_{-3.0}$ & ${0.03367}^{+0.00023}_{-0.00024}$ & ${23.7820}^{+0.0893}_{-0.0890}$\\
        & DESI BAO+$\theta_*^{\textsc{ucs}}$+DESY5 & ${0.3087}^{+0.0075}_{-0.0073}$ & -- & -- & ${4.5}^{+3.1}_{-3.0}$ & ${0.03350}^{+0.00022}_{-0.00023}$ & ${23.8326}^{+0.0068}_{-0.0067}$ \\
    \end{tabular}
    \end{ruledtabular}
\end{table*}

\section{Discussion}\label{sec:discussion}
\subsection{A potential new difference between the early-time and late-time constraints}
We have seen in Figure~\ref{fig:w0-wa-compare} that the UCS analyses of the dynamical dark energy model with CPL parameterization yield constraints that are more consistent with the $\Lambda$CDM model than those from the corresponding full analyses. This shift may hint at some unidentified effects or physical process. However, internal consistency checks of the CMB data have been performed, and no significant inconsistencies have been identified~\citep{Planck-XI-2016}.

Although no substantial discrepancies within the CMB data are currently evident, potential small systematic effects or deviations from the standard early-Universe cosmological model could still influence constraints on late-time cosmology, particularly in the era of precision measurements. While most UCS constraints on cosmological parameters shown in Table~\ref{tab:summary-results} agree with those from the full analyses of corresponding datasets, subtle differences remain. Notably, under the assumption of the $\Lambda$CDM model post-recombination, the UCS constraint on $\Omega_{\rm m}$ from DESI BAO+$\theta_*^{\textsc{ucs}}$ (without SNe Ia) is:
\begin{equation}\label{eq:Om-UCS-LCDM}
\Omega_{\rm m}^{\textsc{ucs}} = 0.2960 \pm 0.0049\,,
\end{equation}
which is lower than the value from the full analysis \citep{DESI:2025zgx}:
\begin{equation}\label{eq:Om-full-LCDM}
\Omega_{\rm m}^{\textsc{full}} = 0.3027 \pm 0.0036\,.
\end{equation}
The full analysis incorporates all observational information, including that used in the UCS analysis. This means the additional pre-recombination information pulls the inferred $\Omega_{\rm m}$ upward.

If we naively treat the information contained in the UCS analysis as statistically independent from the remaining information of the full analysis, we can infer that those other information constrain 
\begin{equation}
    \Omega_{\rm m}^{\rm other} \approx 0.3106 \pm 0.0053\,,
\end{equation}
which deviates from the UCS result by approximately $2\sigma$, despite the mean values differing by only about 2\%. This rough estimate highlights a potential and interesting difference between purely pre-recombination and post-recombination constraints: The former are governed by early-Universe physics and are sensitive to the amplitude of the CMB power spectrum, whereas the latter are predominantly geometric, relying solely on the tight correlation between the sound horizon at recombination and at the drag epoch.

The fact that the UCS constraints on the CPL DE model (removing the pre-recombination information) align more consistently with the $\Lambda$CDM case suggests that the aforementioned difference, if real, may be caused by some unidentified pre-recombination physical processes. However, unlike early-Universe physics proposed to address the Hubble tension, such as Early Dark Energy \citep{Poulin:2018cxd} and interacting neutrinos \citep{Kreisch:2019yzn}, resolving this  difference likely requires some ``opposite'' effect: while those earlier models tend to increase the inferred dark matter density \citep{Hill:2020osr,Jedamzik:2020zmd,Vagnozzi:2021gjh}, addressing the current tension would require lowering it, as seen from the comparison between $\Omega_{\rm m}^{\rm other}$ and $\Omega_{\rm m}^{\textsc{ucs}}$. Physical mechanisms that modulate the scale dependence of the CMB power spectrum amplitude may offer a more promising direction. Indeed, this scale dependence constrains the matter density $\Omega_{\rm m}h^2$ \citep{2018-Planck-cosmo-params, ACT:2025fju}. Interestingly, given the anti-correlation between $\Omega_{\rm m}$ and $H_0$, and the positive correlation between $\Omega_{\rm m}$ and $\sigma_8$ from the CMB constraints \citep{2018-Planck-cosmo-params}, a lower $\Omega_{\rm m}$ would lead to a slightly higher $H_0$ and a slightly lower $\sigma_8$ inferred from CMB \citep{Pedrotti:2024kpn}. This would, in turn, also potentially alleviate both the debated Hubble tension and the $\sigma_8$ tension.\footnote{We also note that local measurements of the Hubble constant remain controversial \citep{Freedman:2023jcz}, and there have been important developments regarding the $\sigma_8$ tension \citep{Wright:2025xka,Chiu:2022qgb}.}

\subsection{The remaining late-time inconsistency}
While the UCS constraint on the CPL DE model is more consistent with the $\Lambda$CDM scenario, a mild deviation at the $\sim$2$\sigma$ level—or slightly higher—persists. However, this discrepancy cannot be clearly attributed to a post-recombination deviation from $\Lambda$CDM. An alternative interpretation is a mismatch between \textit{uncalibrated} standard ruler measurements (DESI BAO+$\theta_*^{\textsc{ucs}}$) and \textit{uncalibrated} standard candle measurements (SNe Ia).\footnote{We stress that both cosmic standards are uncalibrated. Thus, the tension arises from their measured relative evolution of distance with redshift, not from their differences in absolute distance.} Notably, as shown in the second panel of Figure~\ref{fig:w0-wa-compare}, the CPL model constraint based on the earlier Pantheon SNe Ia compilation aligns fully with $\Lambda$CDM, though with somewhat weaker constraining power. This does not imply the earlier Pantheon compilation is more reliable—particularly in light of the larger sample size and improved treatment of systematics in the more recent PanPlus compilation and the other two compilations. Nevertheless, this result suggests that addressing the observed tension between uncalibrated standard rulers and candles—whether through improved control of systematics or via new physics related to SNe Ia—may be another promising approach than invoking dynamical dark energy to explain the remaining late-time inconsistency.

\begin{figure}
    \centering
    \includegraphics[width=0.85\linewidth]{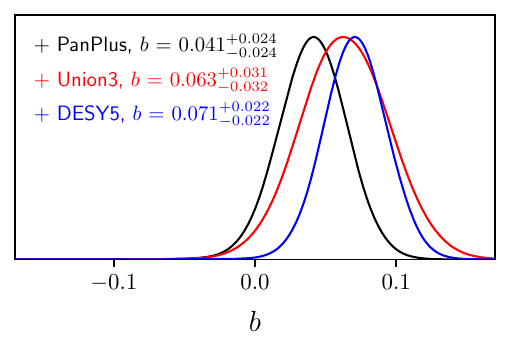}
    \includegraphics[width=0.8\linewidth]{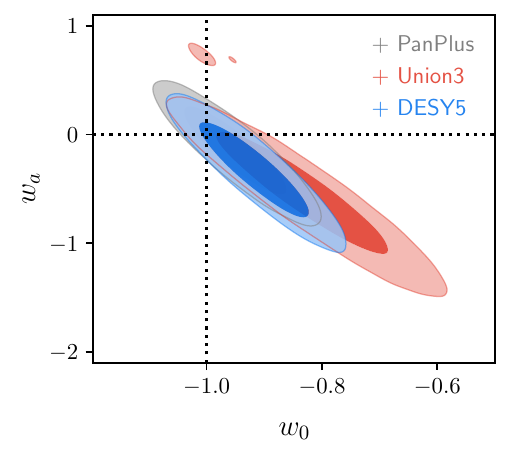}
    \caption{Testing the hypothesis of a redshift-dependent magnitude bias in SNe Ia. Top panel: The marginalized posterior distribution of the magnitude bias parameter $b$, derived using a UCS analysis. The analysis allows $b$ to vary freely within the standard $\Lambda$CDM model. Bottom panel: UCS constraints on the CPL DE model, where $b$ is fixed to the mean value obtained in the top panel. The resultant constraints align well with the $\Lambda$CDM case. }
    \label{fig:test-bias}
\end{figure}

To examine the source of this residual tension, we introduce a redshift-dependent magnitude bias in the SNe Ia data, parameterized as 
\begin{equation}\label{eq:SNIa-bias-redshift}
m^{\rm true}_{\rm B} = m^{\rm obs}_{\rm B} + bz\,,
\end{equation}
A positive $b$ indicates that SNe appear brighter than expected at high redshift, which could be attributed to either systematics or physical effects such as photon attenuation or deviations from the distance duality relation \citep{Keil:2025ysb}. In such a case, the term $bz$ in Eq.~\eqref{eq:SNIa-bias-redshift} can be moved to the right of Eq.~\eqref{eq:magnitude-redshift} and interpreted as a modification to the luminosity distance by combining it with $(1+z)D_{\rm M}$. Similar effects can arise if $\Omega_{\rm m}$ evolves across cosmic time \citep{Colgain:2022nlb,Colgain:2022rxy,Colgain:2024mtg,Colgain:2024ksa}.

We first relax the $b$ parameter within the standard $\Lambda$CDM model and perform parameter inference. The marginalized posterior distribution of $b$ is shown in the top panel of Figure~\ref{fig:test-bias}. All inferred values of $b$ are consistently greater than zero, with the mean ranging from 0.04 to 0.07. This implies that the brightness of SNe Ia at, for example, $z = 1$ would be overestimated by approximately $4\%\sim7\%$. 

Next, we fix the parameter $b$ to its mean values obtained from the above analysis for each data combination and perform parameter inference within the CPL DE model. This approach tests whether such an adjustment can reconcile the derived constraints with the $\Lambda$CDM case. As shown in the bottom panel of Figure~\ref{fig:test-bias}, the resulting parameter constraints are in good agreement with the $\Lambda$CDM results. This suggests that a redshift-dependent bias in the magnitude of SNe Ia could provide an alternative explanation for the remaining late-time discrepancy between the DESI BAO+$\theta_*^{\textsc{ucs}}$ and the SNe Ia data, rather than invoking a dynamical DE component.

\section{Adapting to other late-time cosmological models}
The UCS framework is readily adaptable to a wide range of cosmological analyses beyond the specific datasets and models examined here, such as local voids \citep{Ding:2019mmw,Tomonaga:2023dnz,Banik:2025dlo}, evolving dark matter models \citep{Chen:2025wwn,Kumar:2025etf},  modifications to gravity \citep{Ishak:2018his,Yang:2024kdo,Li:2025cxn}, alternative dark energy models \citep{Joyce:2014kja,Ramadan:2024kmn,Lin:2025gne,Csillag:2025gnz}, and interacting dark sectors \citep{Wang:2016lxa,Vattis:2019efj}. In each of these cases, the primary modification required is to the normalized comoving distance as a function of redshift in Eq.~\eqref{eq:fM-z}. It is also compatible with other low-redshift probes, including cosmic chronometers \citep{Jimenez-2002-chronometers,2017-Moresco-Marulli-CC} and the Alcock-Paczynski test \citep{Dong:2023jtk}, making it a versatile tool for isolating post-recombination physics across multiple observational channels.

\section{Conclusion}
We have developed an analysis framework to test post-recombination cosmological models using uncalibrated standard rulers and candles. This framework, referred to as the \textit{Uncalibrated Cosmic Standards} (UCS) analysis, treats both classes of observables without assuming their absolute calibration. It incorporates the angular size of the sound horizon at recombination in a way that is insensitive to pre-recombination physics and robust against potential systematics in the CMB power spectrum amplitude. The UCS framework operates under minimal assumptions: beyond the constancy of intrinsic source properties, it relies only on the tight correlation between the sound horizon at recombination and at the drag epoch. Despite a modest reduction in constraining power compared to full CMB-based analyses, UCS retains the most essential information and provides a robust, efficient approach to testing post-recombination cosmologies.

We performed a UCS analysis on several commonly adopted late-time cosmological models, although the methodology can be readily extended to other beyond-the-standard cosmological scenarios. Particular attention was given to the dynamical DE model with CPL parameterization. We find that the UCS results are more consistent with the $\Lambda$CDM model compared to those from the full analysis. If such a shift is physical, it is more likely to result from the omission of information concerning the scale dependence of the CMB power spectrum amplitude, which may be influenced by certain pre-recombination physical processes or by potential subtle, undiagnosed systematic effects.

Additionally, we identify a mild inconsistency between uncalibrated standard rulers and uncalibrated SNe Ia data, which contributes to the residual discrepancy between the CPL DE constraints and the $\Lambda$CDM case. Introducing a redshift-dependent magnitude bias in the SNe Ia data effectively resolves this remaining late-time tension, offering an alternative explanation that does not require invoking a dynamical DE component.

A caveat of our work is that we treat the standard $\Lambda$CDM model as a reference case to identify alternative explanations for the recently reported detection of dynamical DE. Nonetheless, the alternatives explored in this study—such as pre-recombination systematics and redshift-dependent biases in SNe Ia magnitudes—demonstrate that apparent signatures of dynamical DE can also arise from systematic effects and early-Universe effects. This underscores the importance of critically evaluating late-time cosmological tensions before attributing them to new physics.

Looking ahead, precision cosmology will be driven by upcoming galaxy surveys with unprecedented statistical power \citep{LSST:2008ijt,Euclid:2024yrr,Zhan2021}. As the statistical uncertainties shrink, the interpretation of cosmological data will become increasingly limited by systematic effects and model assumptions, particularly those tied to early-Universe physics. In this context, the UCS framework provides a robust approach to isolate and test post-recombination cosmology. By disentangling late-time observables from pre-recombination assumptions, UCS stands out as a promising tool for probing the dark sector in the high-precision era.

\begin{acknowledgements}
We thank Xingang Chen, Luca Visinelli and Eoin Ó Colgáin for useful discussions. W.L. acknowledges that this work is supported by the ``Science \& Technology Champion Project" (202005AB160002), the ``Top Team Project" (202305AT350002) and the ``Innovation Team Project'' (202105AE160021), all funded by the ``Yunnan Revitalization Talent Support Program." This work is also supported by the "Key Laboratory of Survey Science of Yunnan Province" (202449CE340002), the ``Yunnan General Grant'' (202401AT070489) and the National Key R\&D Program of China (2024YFA1611600).
\end{acknowledgements}

\bibliography{UCstds}
\bibliographystyle{aasjournal}


\end{document}